\documentclass[twocolumn]{aastex631}

\usepackage{CJK}
\usepackage{amsmath} 

\usepackage{xcolor} 
\definecolor{bg-green}{rgb}{0.8588,0.9333,0.8666}
\definecolor{Q-color}{rgb}{0.5,0.1,0.5}

\shorttitle{Cold Gas in Radiatively Cooling Outflows}
\shortauthors{Qiu et al.}

\received{June 5, 2021}
\revised{July 17, 2021}
\accepted{July 20, 2021}

\submitjournal{\apjl}

\begin{document} 
\begin{CJK*}{UTF8}{gbsn} 

\title{Dynamics and Morphology of Cold Gas in Fast, Radiatively Cooling Outflows:\\ Constraining AGN Energetics with Horseshoes}

\correspondingauthor{Yu Qiu (邱宇)}
\email{yuqiu@pku.edu.cn}

\author[0000-0002-6164-8463]{Yu Qiu}
\affiliation{Kavli Institute for Astronomy and Astrophysics, Peking University, 5 Yiheyuan Road, Haidian District, Beijing, 100871, PRC}

\author[0000-0003-3143-3995]{Haojie Hu}
\affiliation{Kavli Institute for Astronomy and Astrophysics, Peking University, 5 Yiheyuan Road, Haidian District, Beijing, 100871, PRC}
\affiliation{Department of Astronomy, School of Physics, Peking University, 5 Yiheyuan Road, Haidian District, Beijing 100871, PRC}

\author[0000-0001-9840-4959]{Kohei Inayoshi}
\affiliation{Kavli Institute for Astronomy and Astrophysics, Peking University, 5 Yiheyuan Road, Haidian District, Beijing, 100871, PRC}

\author[0000-0001-6947-5846]{Luis C. Ho}
\affiliation{Kavli Institute for Astronomy and Astrophysics, Peking University, 5 Yiheyuan Road, Haidian District, Beijing, 100871, PRC}
\affiliation{Department of Astronomy, School of Physics, Peking University, 5 Yiheyuan Road, Haidian District, Beijing 100871, PRC}

\author[0000-0002-7835-7814]{Tamara Bogdanovi\'c}
\affiliation{Center for Relativistic Astrophysics, School of Physics, Georgia Institute of Technology, 837 State Street, Atlanta, GA 30332, USA}

\author[0000-0002-2622-2627]{Brian R. McNamara}
\affiliation{Department of Physics and Astronomy, University of Waterloo, 200 University Avenue West, Waterloo, ON, N2L 3G1, Canada}
\affiliation{Waterloo Center for Astrophysics, University of Waterloo, 200 University Avenue West, Waterloo, ON, N2L 3G1, Canada}
\affiliation{Perimeter Institute for Theoretical Physics, Waterloo, ON, N2L 2Y5, Canada}


\begin{abstract}
Warm ionized and cold neutral outflows with velocities exceeding $100\,{\rm\\km\,s}^{-1}$ are commonly observed in galaxies and clusters. Theoretical studies however indicate that ram pressure from a hot wind, driven either by the central active galactic nucleus (AGN) or a starburst, cannot accelerate existing cold gas to such high speeds without destroying it. In this work we explore a different scenario, where cold gas forms in a fast, radiatively cooling outflow with temperature $T\lesssim 10^7\,{\rm\\K}$. Using 3D hydrodynamic simulations, we demonstrate that cold gas continuously fragments out of the cooling outflow, forming elongated filamentary structures extending tens of kiloparsecs. For a range of physically relevant temperature and velocity configurations, a ring of cold gas perpendicular to the direction of motion forms in the outflow. This naturally explains the formation of transverse cold gas filaments such as the blue loop and the horseshoe filament in the Perseus cluster. Based on our results, we estimate that the AGN outburst responsible for the formation of these two features drove bipolar outflows with velocity $>2,000\,{\rm\\km\,s}^{-1}$ and total kinetic energy $>8\times10^{57}\,{\rm\\erg}$ about $\sim10$\,Myr ago. We also examine the continuous cooling in the mixing layer between hot and cold gas, and find that radiative cooling only accounts for $\sim10\%$ of the total mass cooling rate, indicating that observations of soft X-ray and FUV emission may significantly underestimate the growth of cold gas in the cooling flow of galaxy clusters.
\end{abstract}


\section{Introduction} \label{sec:intro}
Galactic winds and outflows are commonly observed in and around active galaxies \citep[see][for a review]{Veilleux2005}. Driven either by starbursts or accreting supermassive black holes (SMBHs), the velocity of such outflows ranges from $\lesssim100\,{\rm\\km\,s}^{-1}$ to $\gtrsim1,000\,{\rm\\km\,s}^{-1}$ \citep[e.g.,][]{Filippenko1992}, sometimes exceeding the sound speed of the outflowing material. The short destruction timescale of such outflows by strong shocks raises questions about the origin of the fast-traveling gas, as well as its subsequent evolution and interaction with the ambient medium. Two main branches of models have been proposed to explain the existence of the fast outflows: 

(1) On the one hand, cold gas may be entrained into a fast-moving hot wind \citep[the so-called cloud-crushing problem, e.g.,][]{Klein1994,Cooper2009,Fujita2009,Scannapieco2015,Gronke2018,Sparre2019,Li2020,Kanjilal2021}. However, this mechanism does not seem able to accelerate the bulk of the cold gas to velocities $>100\,{\rm\\km\,s}^{-1}$ due to rapid shock destruction \citep{Zhang2017}, hindering its application to fast outflows. 

(2) On the other hand, cold gas may form within an initially hot, radiatively cooling outflow, which retains high speeds beyond $\sim100\,{\rm\\km\,s}^{-1}$\citep{Wang1995,Thompson2016,Schneider2018,Qiu2020} {and survives longer due to lower relative velocity with the original hot gas.} 

In this work we perform the first dedicated set of simulations to study the thermodynamical and morphological evolution of the cold gas that forms in a fast, initially hot, radiatively cooling outflow with short cooling time ($<10\,{\rm Myr}$). In Section~\ref{sec:sim} we describe the simulation setup and the relevant observational constraints. Section~\ref{sec:result} summarizes the dynamical evolution of the cold gas as well as the accompanying emission features. We then discuss the implications and conclude in Section~\ref{sec:discussion}.


\section{Simulations of Radiatively Cooling Outflows}\label{sec:sim}

In order to facilitate the comparison with observations and demonstrate the applicability of the model, we initialize the simulations inside the deep potential well of a galaxy cluster, where both gravity and ram pressure from the ambient intracluster medium (ICM) play important roles in the dynamical evolution of the outflowing gas. Observationally, \citet{Russell2019} have measured $10^9-10^{11}\,M_\sun$ molecular gas in central cluster galaxies using ALMA data. Most of the molecular gas lies in radial filamentary structures extending up to tens of kiloparsecs, with line-of-sight velocities below $\sim500\,{\rm\\km\,s}^{-1}$. The relatively low velocity and the wide-spread radial geometry have led to the postulation that the filaments are entrained by rising radio bubbles inflated by AGN jets \citep[e.g.,][]{Revaz2008,McNamara2016}. Simulations performed by \citet{Duan2018} however find that the trailing cold gas is confined to the cluster core by gravity, unable to form extended filaments. Another challenge facing past AGN feedback models is the presence of dust \citep[e.g.,][]{Mittal2012}, which cannot survive thermal sputtering \citep{Draine1979} if the cold gas forms due to thermal instabilities of the ambient hot ICM \citep[e.g.,][]{McCourt2012, Gaspari2012, Li2014}.

Using cluster-scale simulations that launch multiphase outflows from the central AGN \citep[see][and references therein for a detailed comparison of AGN feedback models]{Qiu2018}, \citet{Qiu2020} proposed that the cold gas forms in radiatively cooling outflows with initial temperature $T\lesssim10^7\,{\rm\\K}$. Because the cooling time ($\lesssim10\,{\rm\\Myr}$) is shorter than the dust sputtering timescale ($\sim10$\,Myr), and the timescale for the gas to reach maximum height ($\sim100\,{\rm\\Myr}$), the gas fragments into dusty cold clumps along the rising trajectory. Additionally, due to ram pressure from the intfracluster plasma, the outflow speed is significantly decelerated before fragmentation, yielding cold gas velocities lower than predictions based on simple, ballistic estimates. Filaments formed in this way share properties similar to those observed in Perseus and other cool-core clusters \citep[e.g.,][]{Conselice2001,Qiu2019, Qiu2018}.

\begin{deluxetable}{lrccc}[t!]
\tablecaption{Simulation Parameters}
\tablewidth{0pt}
\tablehead{
\colhead{Simulation} 	& \colhead{Res.} 	& \colhead{$v_{\rm out}$} 				& \colhead{$T_{\rm out}$} 	& \colhead{$M_{\rm out}$} \\
\colhead{ID} 			& \colhead{(pc)} 	& \colhead{(${\rm km\,s}^{-1}$)} 	& \colhead{$(10^7\,{\rm K}$)} 				& \colhead{($M_\sun$)}
}
\startdata
LTlr		& 120	& 1,200		& 1.0	 	& $10^8$ 	\\
LTmr		& 60		& 1,200		& 1.0 	& $10^8$ 	\\
LThr		& 30		& 1,200		& 1.0		& $10^8$ 	\\
HTlr		& 120	& 2,000		& 1.2		& $10^8$ 	\\
HTmr	& 60		& 2,000		& 1.2		& $10^8$ 	\\
HThr		& 30		& 2,000		& 1.2		& $10^8$ 	\\
\enddata
\label{tab:params}
\tablecomments{Simulation ID indicates the initial temperature of the outflowing clump, as well as the simulation resolution: low temperature (LT), high temperature (HT), low, medium, and high resolution (lr, mr, hr).}
\end{deluxetable}

Following the analytic model laid out in \citet{Qiu2020}, we perform a series of 3D high-resolution simulations of a radiatively cooling outflow traveling radially in a cluster potential using the code ENZO \citep{Bryan2014}. The cluster is modeled after Perseus and initialized with hot plasma in hydrostatic equilibrium with the background gravity \citep[contributed by the dark matter halo, the central galaxy, and the SMBH, as detailed in][]{Qiu2018}. The outflowing gas, with temperature $T_{\rm\\out}$ and metallicity $Z=0.02$, is initially set up as a spherical clump in thermal pressure equilibrium with the surrounding ICM whose core temperature $T\approx3\times10^7\,{\rm\\K}$, {electron number density $n_e=0.07\,{\rm\\cm}^{-3}$}, and a lower $Z=0.011$. The elevated metallicity of the outflowing gas allows us to model the transport of metals by the outflows \citep[e.g.,][]{Kirkpatrick2015} and serves as a natural tracer for the outflowing gas. {It also slightly increases the cooling rate of the outflow (by a factor between $1-2$ depending on the gas temperature), but does not alter our results significantly.} We fix the clump mass $M_{\rm\\out}=10^8\,M_\sun$, in agreement with typical molecular gas clump measurements in Perseus \citep[e.g.,][]{Salome2006,Salome2008,Salome2011}. The center of the clump is placed 1\,kpc away from the SMBH along the $x$-axis. We then give the gas clump an initial outward velocity $v_{\rm\\out}$, and follow its subsequent evolution for 40\,Myr, which is roughly four times the cooling timescale.

\begin{figure*}[t!]
\centering
\includegraphics[scale=1]{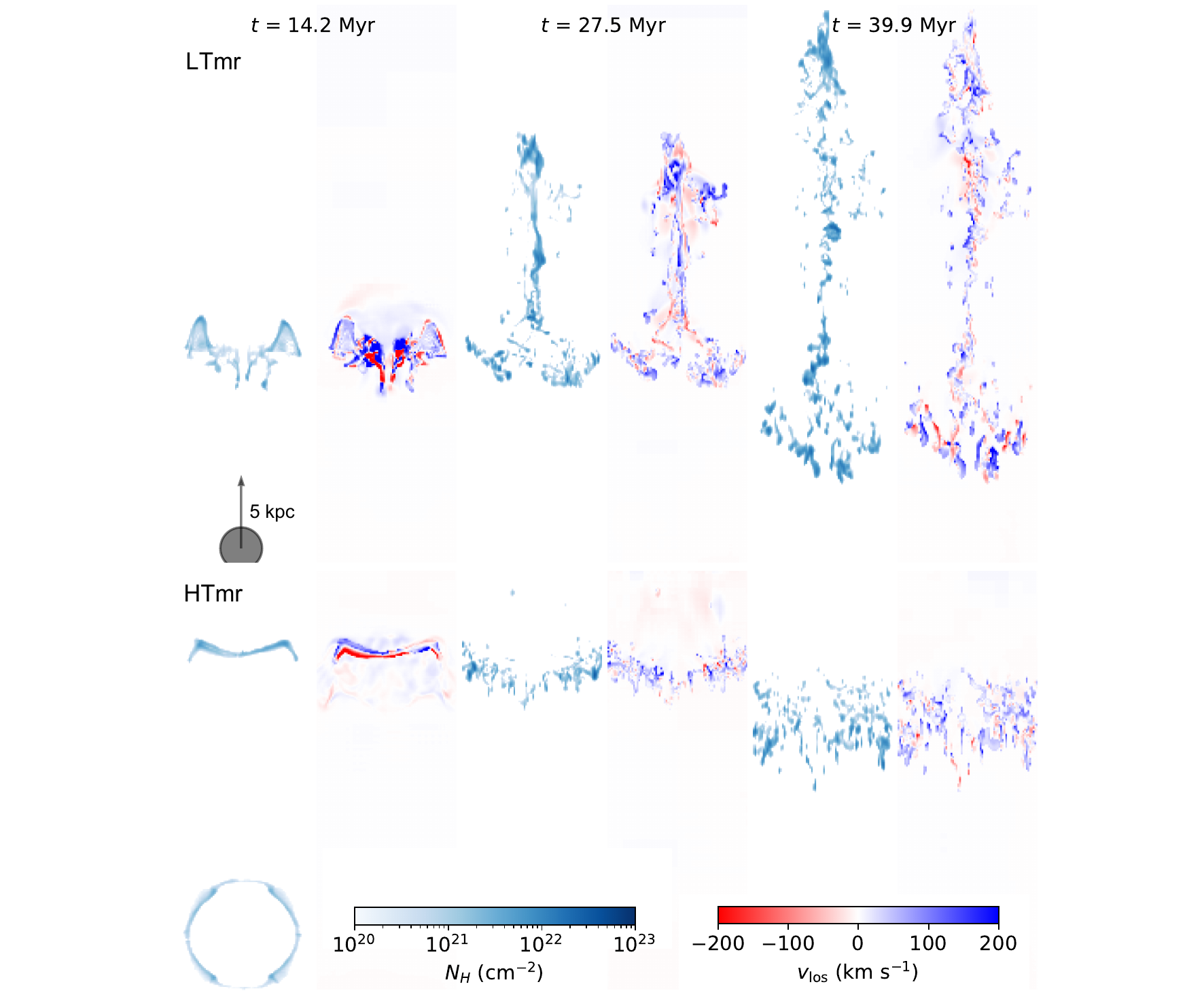}
\caption{Spatial and velocity distribution of the cold gas that forms in the outflows, shown as projected neutral hydrogen column density, and mass-weighted line-of-sight velocity at three snapshots in the simulations LTmr and HTmr. Outflows are launched from the bottom of the panels with initial velocities upwards (shown by the grey circle and arrow). Projection along the flow direction that displays a ring structure is shown in the inset of the bottom left panel.}
\label{fig:image}
\end{figure*}

We select the initial outflow velocity $v_{\rm\\out}$ based on the constraint that when cold gas fragments, the velocity is below $500\,{\rm\\km\,s}^{-1}$ (see model predictions in Figure~\ref{fig:model}). $v_{\rm\\out}$ is uniform within the clump, and aligned with the $x$-axis. As summarized in Table~\ref{tab:params}, we select two initial temperatures in order to investigate the different cold gas morphologies that can arise from the outflows, and note that $T_{\rm\\out}\leq10^7\,{\rm\\K}$ outflows share similar filamentary structure, while $T_{\rm\\out}>1.3\times10^7\,{\rm\\K}$ leads to net heating of the outflow, without forming cold gas in our simulations. {The sound speed for $\sim10^7\,{\rm\\K}$ plasma is $\sim500\,{\rm\\km\,s}^{-1}$, which means the outflows explored in our simulations have Mach number of $2-4$. Therefore, heating of the outflow is provided by a combination of shock-heating and mixing with the entrained hot ICM. $T\approx1.2\times10^7\,{\rm\\K}$ represents the balance point between heating and radiative cooling, which leads to partial cold gas formation, as described in Section~\ref{sec:result}}. 

In order to explore the impact of simulation resolution, we perform 3 simulations with increasingly more refinement levels for each outflow temperature choice. The criteria for the adaptive mesh refinement include density, cooling time, Jeans length, and shocks \citep[see][for more information]{Bryan2014, Qiu2018}. {We model self-gravity of the gas in our simulations.} This allows us to capture the shock-heated, radiatively cooling, and gravitationally collapsing gas in detail, but we note that the Jeans length of the cold, dense gas after it forms is not always resolved at the resolutions in our simulations \citep{Truelove1997}. 


\section{Dynamical Evolution of Cold Gas}  \label{sec:result}

In this section we present the simulation results of the radiatively cooling outflow as it decelerates by gravity and ram pressure, gradually loses thermal energy, and fragments into filaments of cold gas. In order to assess the simple analytic model laid out in \citet{Qiu2020}, we also calculate and trace properties of the center of mass for the original outflowing gas in Section~\ref{sec:velocity}. 

\subsection{Spatial Distribution} \label{sec:spatial}

We first examine the spatial distribution of the cold gas after it fragments out of the radiatively cooling plasma. With $T_{\rm\\out}\approx10^7\,{\rm\\K}$ in the core of the simulated cluster, the cooling time of the outflow is $\sim10\,{\rm\\Myr}$. In Figure~\ref{fig:image} we show the projected neutral hydrogen column density, as well as the density-weighted line-of sight velocity for three snapshots between $10-40\,{\rm\\Myr}$ for the representative, medium-resolution runs, LTmr and HTmr. In both cases, the cold gas first appears at $x=10-20\,{\rm\\kpc}$, {with hydrogen column densities $N_H\sim10^{21}-10^{22}\,{\rm\\cm}^{-2}$}. Before cooling to the neutral phase, the clump is significantly decelerated by the ram pressure from the core plasma, and the cold gas velocity along the direction of motion, $v_x$, is below $500\,{\rm\\km\,s}^{-1}$. Subsequently, the gas fragments further by its self-gravity into smaller cold gas clumps. Throughout the 40-Myr evolution, the bulk of the cold gas never attained speeds higher than $500\,{\rm\\km\,s}^{-1}$, in agreement with observational constraints \citep[e.g.,][]{GM2018}\footnote{The SITELLE observation of the warm, ionized filaments emitting H$\alpha$ photons in the Perseus cluster \citep{GM2018} reveals a rich velocity structure, including a low velocity dispersion and a smooth velocity gradient along the length of the filaments. For brevity we defer the detailed comparison of the structures to a future work.}.

The morphologies of the cold gas clumps are however different in these two cases. In LTmr, after the first generation of cold gas appears at the {interface between the outflow and the ICM, $\sim15\,{\rm\\kpc}$ from the center, there is continuous cold gas formation in the remaining ionized outflow that rises to 40\,kpc}. By the end of the 40-Myr simulation, different generations of cold gas span over 10\,kpc in length, with both positive and negative $x$-velocities. However, in HTmr with a slightly higher initial temperature, cold gas first appears as a ring on the outer edge of the outflowing clump due to increased density and elevated cooling \citep[analogous to vortex rings in a turbulent fluid, e.g., the radio torus in M87,][]{Owen2000,Churazov2001,Forman2017}. The gas interior to the cold ring eventually penetrates through, but due to heating and mixing with the entrained ICM, there is very little subsequent cold gas formation. Comparing these two simulations, the morphology of the cold gas depends sensitively on the initial condition of the outflow.

\subsection{Comparison with the Analytic Model} \label{sec:velocity}

\begin{figure*}[t!]
\centering
\includegraphics[scale=1, trim=0mm 8mm 0mm 0mm, clip=true]{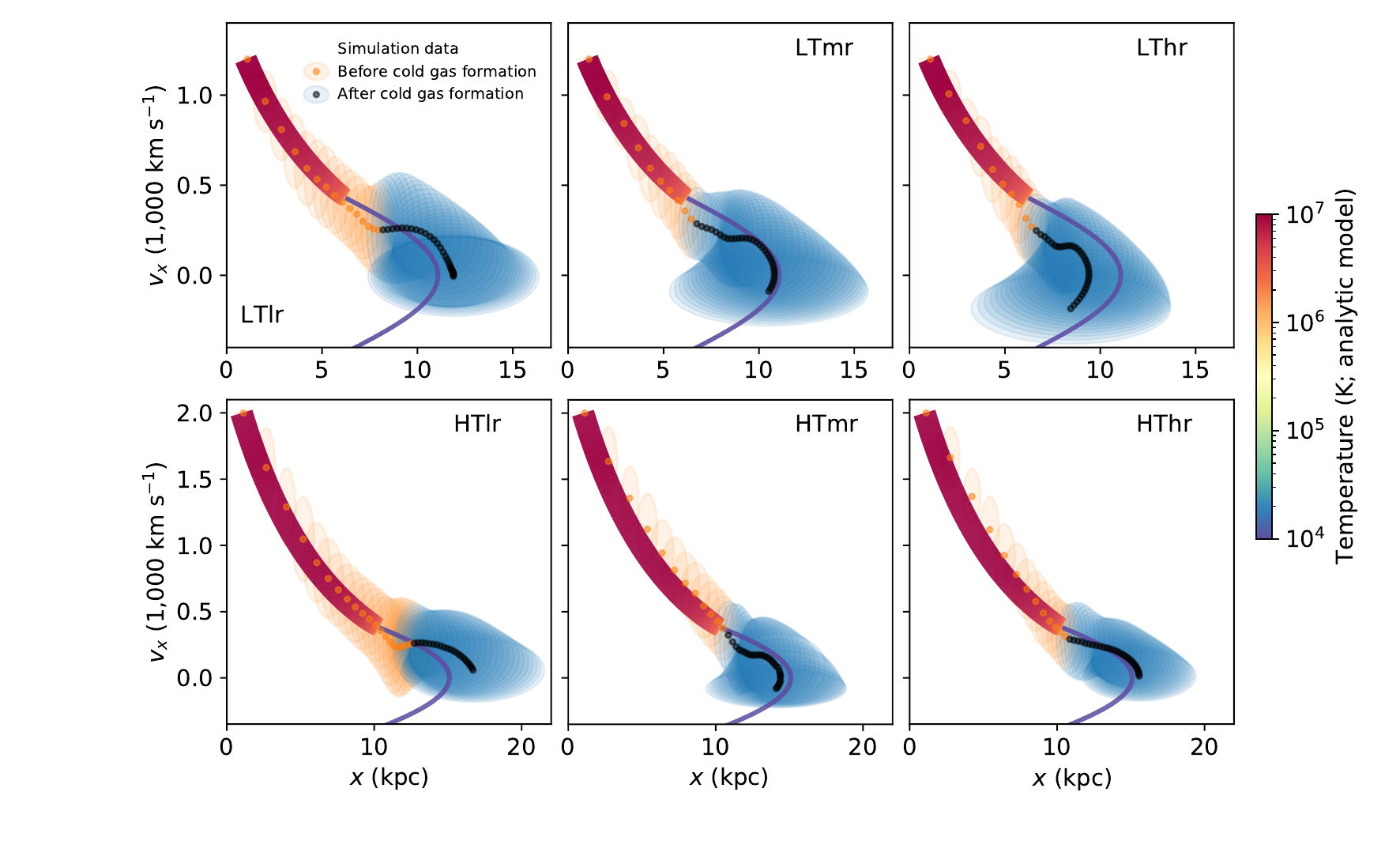}
\caption{$v_x$ evolution as a function of $x$-location of the outflowing gas traced by the elevated metallicity, sampled every $\approx1\,{\rm\\Myr}$. Orange and black circles represent the location and velocity of the center of mass of the outflow. The color changes after cold gas forms in the outflow. Orange and blue ellipses centered on each data point represent the standard deviation of the properties of the traced gas. Data points are overlaid on top of the analytic model from \citet{Qiu2020}. The color and width of the lines represent the temperature and relative size of the outflowing clump.}
\label{fig:model}
\end{figure*}

The fragmentation and evolution of individual cold gas clumps are difficult to predict, but their collective behavior can be extracted for comparison with the analytic model. In Figure~\ref{fig:model} we plot the evolution of $v_x$ as a function of $x$-location for the center of mass of the initial outflowing gas, overlaid on top of the model prediction proposed in \citet{Qiu2020} for each simulation. {The analytic model assumes a sphere of outflowing clump is traveling in the cluster potential, and in thermal pressure equilibrium with the ICM. In each timestep it calculates the radiative cooling rate \citep{Schure2009}, the deceleration by gravity and ram pressure from the ICM, and updates the temperature, velocity, location, and size (cross section) of the outflowing clump.} The original outflow in the simulations is traced by the elevated metal content in the gas cells. Each data point, sampled every $\sim1\,{\rm\\Myr}$, is surrounded by an ellipse, whose vertical and horizontal axes represent the mass-weighted standard deviation of the gas velocity and position. This allows us to visualize the span of the individual cold gas clumps in both real and velocity space. The color of the ellipse changes from orange to blue after cold gas forms in the outflow.

Overall the evolution of the outflowing gas in the $(x,v_x)$ diagram agrees with the analytic model. In the first $\sim10\,{\rm\\Myr}$, the outflow cools radiatively and is decelerated to speeds $<500\,{\rm\\km\,s}^{-1}$. After cold gas forms, effects from the ram pressure diminishes, and the subsequent evolution can be described by a ballistic trajectory. {Due to the faster initial speed in HT simulations, the center of mass reaches 15\,kpc away from the cluster center, further than LT simulations. In both cases, the standard deviation of the gas velocity \citep[$\lesssim200\,{\rm\\km\,s}^{-1}$, in agreement with][]{GM2018} is much lower compared with the launching speed ($>1,000\,{\rm\\km\,s}^{-1}$)}.

Apart from physical processes, simulation resolution also alters the distribution of the gas, but to a lesser degree. In both LT and HT cases, higher resolution leads to higher gas density that boosts the local radiative cooling rate of the ionized plasma. This results in earlier cold gas formation in medium and high-resolution simulations compared with low-resolution runs. In addition, gas in low-resolution simulations also tends to travel further, due to less turbulent motions developed on a coarser grid. However, simulation HTmr slightly breaks the trend with a lower height in the cluster atmosphere compared with HThr. As mentioned in Section~\ref{sec:spatial}, in HTmr the cold gas forms a ring, and the rest of the outflow continues to rise and mix with the surrounding ICM. This leads to a center of mass location larger than the spatial reach of the cold gas (as shown in Figure~\ref{fig:image}). The hot and rising outflow is again subject to ram pressure deceleration, thus reducing the ultimate spatial reach of the center of mass. The same ring structure does not appear in HThr, due to boosted radiative cooling with higher densities, leading to a cold gas distribution similar to that shown for LTmr in Figure~\ref{fig:image}. We do not however seek resolution convergence for the gas morphology in this work, because physical processes not explored in our hydrodynamic simulations likely stabilize the cold gas against fragmentation, such as magnetic fields, conduction, and self-shielding \citep[e.g.,][]{Li2020}. HTmr represents a unique case in the explored parameter space, which indicates that the initial conditions and environmental properties must be very stringent for the ring structure to appear. 

\subsection{Radiative Cooling {\it vs.} Mixing} \label{sec:mix}

\begin{figure*}[t!]
\centering
\includegraphics[scale=0.8]{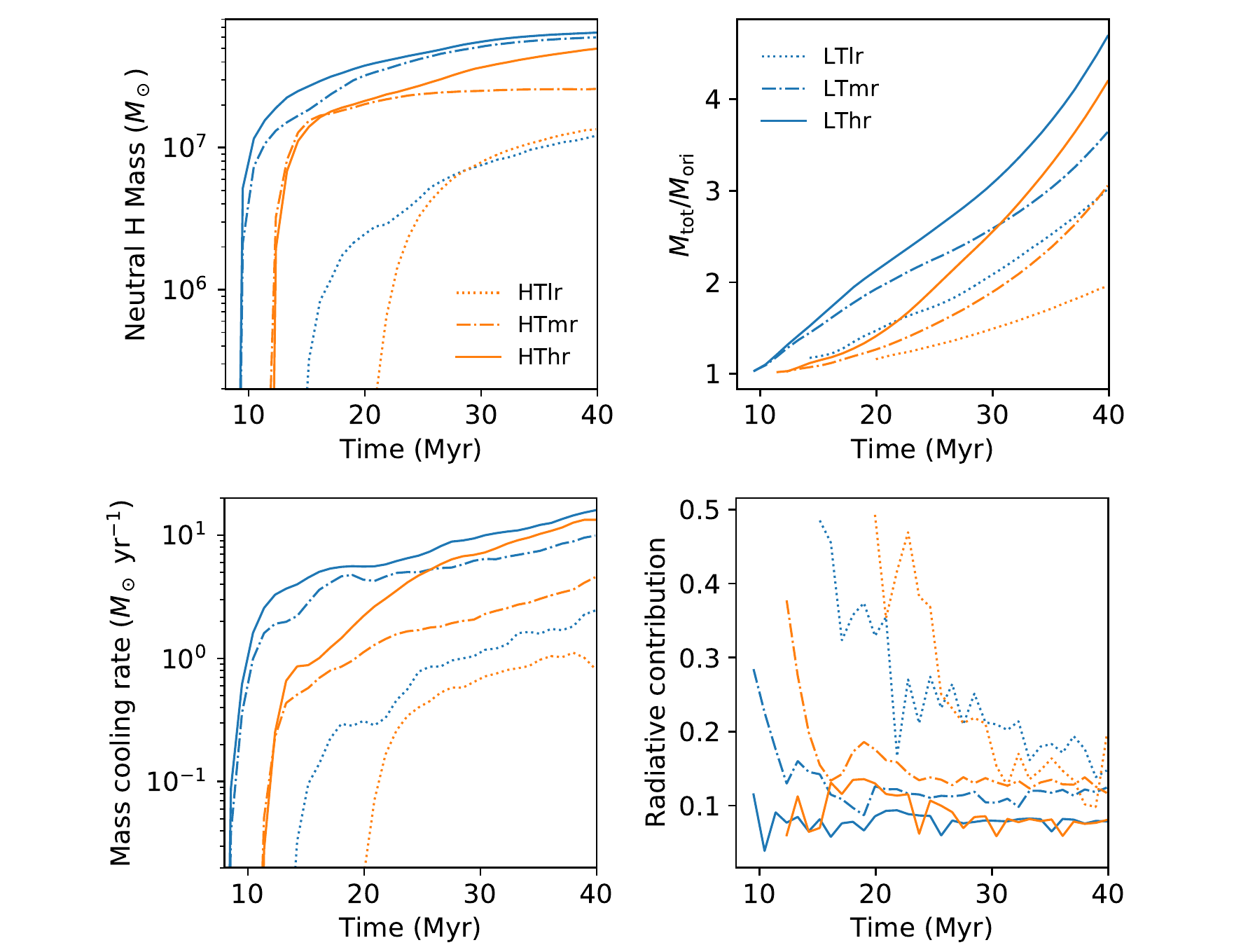}
\caption{Evolution of the cold gas mass and mass cooling rate in simulations. {\it Top left:} Total amount of cold gas in the simulation traced by neutral hydrogen. {\it Top right:} Ratio between total cold gas ($M_{\rm\\tot}$) and the component from the original outflow ($M_{\rm\\ori}$). The increasing ratio indicates additional cold gas formation from the ICM. {\it Bottom left:} Total cold gas formation rate contributed by both radiative cooling and non-radiative mixing. {\it Bottom right:} Fraction of mass cooling rate inferred from the X-ray emissivity of the plasma with temperature between $10^6-10^7\,{\rm\\K}$. Contribution by radiative cooling only accounts for $10-20\%$ of the total mass cooling rate.}
\label{fig:mix}
\end{figure*}

Observationally, the cold gas embedded in the hot ICM is surrounded by a mixing layer of gas with soft X-ray \citep[$\sim0.5\,{\rm\\keV}$, e.g.,][]{Fabian2006} and FUV emission \citep[e.g., O\,{\sc vi} emission from $T\sim10^{5.5}\,{\rm\\K}$ gas,][]{Bregman2006}. Whether the plasma is cooling primarily radiatively or through non-radiative mixing\footnote{Note here ``non-radiative'' refers to the bypass of X-ray and FUV emission. The mixed gas may still emit photons at lower energies.} with the cold gas complicates the interpretation of its emission features and mass estimates. In this subsection we compare the two cooling processes and access their relative significance in our simulations.

The top left panel of Figure~\ref{fig:mix} shows the evolution of the cold gas mass in all 6 simulations, represented by their neutral hydrogen content, ranging between $M_{\rm\\tot}\sim10^6-10^8\,M_\sun$. Due to unresolved dense gas in low-resolution simulations, the final cold gas mass is lower by a factor of a few compared with medium- and high-resolution simulations. Using the elevated metallicity as a tracer, we also identify the component within the cold gas that originates from the initial outflowing gas ($M_{\rm\\ori}$). Shown in the top right panel, the ratio of total-to-original cold gas mass evolves from $\approx1$ when cold gas first appears around 10\,Myr, to $\approx2-4$ towards the end of the simulations, indicating that the cold gas is constantly mixing with the ambient medium. 

In order to separate the two cooling processes that can lead to the continuous growth of cold gas in our simulations, in the bottom panels of Figure~\ref{fig:mix} we compare the additional cold gas formation rate through mixing, $d(M_{\rm\\tot}-M_{\rm\\ori})/dt$, and the radiative cooling rate in the mixing layer between hot and cold gas. The radiative cooling rate is calculated by the thermal X-ray emission from gas with temperature $10^6\,{\rm\\K}<T<10^7\,{\rm\\K}$ using the pyXSIM package \citep{ZuHone2014}, and then converting to mass cooling rates assuming an average plasma temperature of $5\times10^6\,{\rm\\K}$. This reveals that the radiative cooling rate is lower than the mixing cooling rate by a factor close to 10. Applying this ratio to real clusters and boosting the cold gas mass from $10^8\,M_\sun$ to a typical $10^{10}\,M_\sun$, our analysis indicates that while the mass cooling rate inferred by soft X-ray and FUV emission is on the order of $\sim10\,M_\sun\,{\rm\\yr}^{-1}$ \citep{Fabian2006,Bregman2006}, cooling through non-radiative mixing between hot and cold gas can be 10 times faster. We caution, however, the high mixing cooling rate may be due to insufficient simulation resolution, as discussed in \citet{Qiu2021}.


\section{Discussion \& Application}\label{sec:discussion}
In this work we perform the first hydrodynamic simulations to examine the dynamical and morphological evolution of cold gas that cools and fragments in an initially hot, radiatively cooling outflow in a galaxy cluster. Cold gas that forms in this way naturally retains the high speeds of the outflow $>100\,{\rm\\km\,s}^{-1}$, which helps explain the abundance of fast neutral and warm ionized gas observed in galaxies and clusters. We summarize the main results and implications below:

1. Continuous cold gas formation in the fast, radiatively cooling outflow results in a distribution of the gas along the direction of motion, which naturally explains the $\sim10$-kpc-long cold gas filaments commonly observed in cool-core clusters. Due to ram pressure deceleration in the cluster environment, the cold gas velocity is significantly reduced before formation to speeds between $100-500\,{\rm\\km\,s}^{-1}$ \citep{Qiu2020}, in line with observational constraints.

2. {Under certain physically plausible conditions where heating balances radiative cooling}, a ring of cold gas perpendicular to the direction of motion may form out of the outflowing gas. This is similar to the blue loop and the horseshoe filament in the Perseus cluster \citep[e.g.,][]{Fabian2006}, the latter was thought to be lifted by rising X-ray cavities \citep{Fabian2003}. These two features are located on opposite sides of the central AGN, indicating that they may be driven by the same AGN outburst $10-20$\,Myr ago that pushed out $\sim10^8\,M_\sun$ \citep{Salome2008,Salome2011} of radiatively cooling outflow in each direction. Assuming the initial outflow speed is $>2,000\,{\rm\\km\,s}^{-1}$, as found in simulation HTmr, and the duration of the outburst is $\sim1\,{\rm\\Myr}$, this implies a total kinetic energy $>8\times10^{57}\,{\rm\\erg}$, and an average AGN kinetic feedback power $\gtrsim2\times10^{44}\,{\rm\\erg\,s}^{-1}$. Hence, properties of the cold gas filaments in galaxy clusters can be used to probe past AGN activity \citep[see also][]{Qiu2019}.

{3. Adopting the outflow-to-accretion mass-loading factor $m=200-500$ found in \citet{Qiu2021} for AGNs in giant elliptical galaxies, the $2\times10^8\,M_\sun$ outflow estimated above also indicates that the central SMBH accreted at least $4\times10^5\,M_\sun$ during the feeding episode. Assuming 10\% feedback efficiency \citep{Churazov2005}, this yields a total energy output of $\sim10^{59}\,{\rm\\erg}$, primarily in hot outflows exceeding $10^8\,{\rm\\K}$ that are difficult to detect with existing X-ray data \citep{Qiu2021}. Apart from heating the ICM, if a fraction of the cold gas driven by the central AGN collapses and forms stars, such as in the blue loop of the Perseus cluster \citep{Fabian2008,Canning2010}, this provides a channel for SMBHs to positively contribute to the star formation in giant elliptical galaxies, albeit delayed by $\sim10\,{\rm\\Myr}$.}

4. The mixing layer between hot and cold gas contributes significantly to the mass growth rate of the cold gas, tripling the gas content in 30\,Myr. In comparison, the radiative cooling rate in the mixing layer is $\sim10$ times lower. This indicates that the hot plasma may be cooling non-radiatively at a much faster rate than inferred from X-ray and FUV observations in the cooling flow of galaxy clusters, without additional local processes to heat or isolate the cold gas. 

{Our simulations of radiatively cooling outflows assume initial thermal pressure equilibrium between the outflow and the ICM, which fixes the initial density and cooling rate of the outflow for a given temperature. In real clusters, the pressure of the outflows may fluctuate around the equilibrium value and change the radiative cooling rate by a significant factor. Our parameter choice thus represents an intermediate scenario and provides a baseline for connecting AGN activity with the distribution of the cold gas in galaxies and clusters.}


\begin{acknowledgements}
This work is supported by the National Key R\&D Program of China (2016YFA0400702), the National Natural Science Foundation of China (11721303, 11991052, 11950410493, 12003003, 12073003), the China Postdoctoral Science Foundation (2020T130019), and the High-Performance Computing Platform of Peking University.
\end{acknowledgements}



\begin{thebibliography}{}
\expandafter\ifx\csname natexlab\endcsname\relax\def\natexlab#1{#1}\fi
\providecommand{\url}[1]{\href{#1}{#1}}
\providecommand{\dodoi}[1]{doi:~\href{http://doi.org/#1}{\nolinkurl{#1}}}
\providecommand{\doeprint}[1]{\href{http://ascl.net/#1}{\nolinkurl{http://ascl.net/#1}}}
\providecommand{\doarXiv}[1]{\href{https://arxiv.org/abs/#1}{\nolinkurl{https://arxiv.org/abs/#1}}}

\bibitem[{{Bregman} {et~al.}(2006){Bregman}, {Fabian}, {Miller}, \&
  {Irwin}}]{Bregman2006}
{Bregman}, J.~N., {Fabian}, A.~C., {Miller}, E.~D., \& {Irwin}, J.~A. 2006,
  \apj, 642, 746, \dodoi{10.1086/501112}

\bibitem[{{Bryan} {et~al.}(2014){Bryan}, {Norman}, {O'Shea}, {Abel}, {Wise},
  {Turk}, {Reynolds}, {Collins}, {Wang}, {Skillman}, {Smith}, {Harkness},
  {Bordner}, {Kim}, {Kuhlen}, {Xu}, {Goldbaum}, {Hummels}, {Kritsuk}, {Tasker},
  {Skory}, {Simpson}, {Hahn}, {Oishi}, {So}, {Zhao}, {Cen}, {Li}, \& {Enzo
  Collaboration}}]{Bryan2014}
{Bryan}, G.~L., {Norman}, M.~L., {O'Shea}, B.~W., {et~al.} 2014, \apjs, 211,
  19, \dodoi{10.1088/0067-0049/211/2/19}

\bibitem[{{Canning} {et~al.}(2010){Canning}, {Fabian}, {Johnstone}, {Sanders},
  {Conselice}, {Crawford}, {Gallagher}, \& {Zweibel}}]{Canning2010}
{Canning}, R.~E.~A., {Fabian}, A.~C., {Johnstone}, R.~M., {et~al.} 2010,
  \mnras, 405, 115, \dodoi{10.1111/j.1365-2966.2010.16474.x}

\bibitem[{{Churazov} {et~al.}(2001){Churazov}, {Br{\"u}ggen}, {Kaiser},
  {B{\"o}hringer}, \& {Forman}}]{Churazov2001}
{Churazov}, E., {Br{\"u}ggen}, M., {Kaiser}, C.~R., {B{\"o}hringer}, H., \&
  {Forman}, W. 2001, \apj, 554, 261, \dodoi{10.1086/321357}

\bibitem[{{Churazov} {et~al.}(2005){Churazov}, {Sazonov}, {Sunyaev}, {Forman},
  {Jones}, \& {B{\"o}hringer}}]{Churazov2005}
{Churazov}, E., {Sazonov}, S., {Sunyaev}, R., {et~al.} 2005, \mnras, 363, L91,
  \dodoi{10.1111/j.1745-3933.2005.00093.x}

\bibitem[{{Conselice} {et~al.}(2001){Conselice}, {Gallagher}, \&
  {Wyse}}]{Conselice2001}
{Conselice}, C.~J., {Gallagher}, John~S., I., \& {Wyse}, R. F.~G. 2001, \aj,
  122, 2281, \dodoi{10.1086/323534}

\bibitem[{{Cooper} {et~al.}(2009){Cooper}, {Bicknell}, {Sutherland}, \&
  {Bland-Hawthorn}}]{Cooper2009}
{Cooper}, J.~L., {Bicknell}, G.~V., {Sutherland}, R.~S., \& {Bland-Hawthorn},
  J. 2009, \apj, 703, 330, \dodoi{10.1088/0004-637X/703/1/330}

\bibitem[{{Draine} \& {Salpeter}(1979)}]{Draine1979}
{Draine}, B.~T., \& {Salpeter}, E.~E. 1979, \apj, 231, 77,
  \dodoi{10.1086/157165}

\bibitem[{{Duan} \& {Guo}(2018)}]{Duan2018}
{Duan}, X., \& {Guo}, F. 2018, \apj, 861, 106, \dodoi{10.3847/1538-4357/aac9ba}

\bibitem[{{Fabian} {et~al.}(2008){Fabian}, {Johnstone}, {Sanders}, {Conselice},
  {Crawford}, {Gallagher}, \& {Zweibel}}]{Fabian2008}
{Fabian}, A.~C., {Johnstone}, R.~M., {Sanders}, J.~S., {et~al.} 2008, \nat,
  454, 968, \dodoi{10.1038/nature07169}

\bibitem[{{Fabian} {et~al.}(2003){Fabian}, {Sanders}, {Crawford}, {Conselice},
  {Gallagher}, \& {Wyse}}]{Fabian2003}
{Fabian}, A.~C., {Sanders}, J.~S., {Crawford}, C.~S., {et~al.} 2003, \mnras,
  344, L48, \dodoi{10.1046/j.1365-8711.2003.06856.x}

\bibitem[{{Fabian} {et~al.}(2006){Fabian}, {Sanders}, {Taylor}, {Allen},
  {Crawford}, {Johnstone}, \& {Iwasawa}}]{Fabian2006}
{Fabian}, A.~C., {Sanders}, J.~S., {Taylor}, G.~B., {et~al.} 2006, \mnras, 366,
  417, \dodoi{10.1111/j.1365-2966.2005.09896.x}

\bibitem[{{Filippenko} \& {Sargent}(1992)}]{Filippenko1992}
{Filippenko}, A.~V., \& {Sargent}, W. L.~W. 1992, \aj, 103, 28,
  \dodoi{10.1086/116038}

\bibitem[{{Forman} {et~al.}(2017){Forman}, {Churazov}, {Jones}, {Heinz},
  {Kraft}, \& {Vikhlinin}}]{Forman2017}
{Forman}, W., {Churazov}, E., {Jones}, C., {et~al.} 2017, \apj, 844, 122,
  \dodoi{10.3847/1538-4357/aa70e4}

\bibitem[{{Fujita} {et~al.}(2009){Fujita}, {Martin}, {Mac Low}, {New}, \&
  {Weaver}}]{Fujita2009}
{Fujita}, A., {Martin}, C.~L., {Mac Low}, M.-M., {New}, K. C.~B., \& {Weaver},
  R. 2009, \apj, 698, 693, \dodoi{10.1088/0004-637X/698/1/693}

\bibitem[{{Gaspari} {et~al.}(2012){Gaspari}, {Ruszkowski}, \&
  {Sharma}}]{Gaspari2012}
{Gaspari}, M., {Ruszkowski}, M., \& {Sharma}, P. 2012, \apj, 746, 94,
  \dodoi{10.1088/0004-637X/746/1/94}

\bibitem[{{Gendron-Marsolais} {et~al.}(2018){Gendron-Marsolais},
  {Hlavacek-Larrondo}, {Martin}, {Drissen}, {McDonald}, {Fabian}, {Edge},
  {Hamer}, {McNamara}, \& {Morrison}}]{GM2018}
{Gendron-Marsolais}, M., {Hlavacek-Larrondo}, J., {Martin}, T.~B., {et~al.}
  2018, \mnras, 479, L28, \dodoi{10.1093/mnrasl/sly084}

\bibitem[{{Gronke} \& {Oh}(2018)}]{Gronke2018}
{Gronke}, M., \& {Oh}, S.~P. 2018, \mnras, 480, L111,
  \dodoi{10.1093/mnrasl/sly131}

\bibitem[{{Kanjilal} {et~al.}(2021){Kanjilal}, {Dutta}, \&
  {Sharma}}]{Kanjilal2021}
{Kanjilal}, V., {Dutta}, A., \& {Sharma}, P. 2021, \mnras, 501, 1143,
  \dodoi{10.1093/mnras/staa3610}

\bibitem[{{Kirkpatrick} \& {McNamara}(2015)}]{Kirkpatrick2015}
{Kirkpatrick}, C.~C., \& {McNamara}, B.~R. 2015, \mnras, 452, 4361,
  \dodoi{10.1093/mnras/stv1574}

\bibitem[{{Klein} {et~al.}(1994){Klein}, {McKee}, \& {Colella}}]{Klein1994}
{Klein}, R.~I., {McKee}, C.~F., \& {Colella}, P. 1994, \apj, 420, 213,
  \dodoi{10.1086/173554}

\bibitem[{{Li} \& {Bryan}(2014)}]{Li2014}
{Li}, Y., \& {Bryan}, G.~L. 2014, \apj, 789, 153,
  \dodoi{10.1088/0004-637X/789/2/153}

\bibitem[{{Li} {et~al.}(2020){Li}, {Hopkins}, {Squire}, \& {Hummels}}]{Li2020}
{Li}, Z., {Hopkins}, P.~F., {Squire}, J., \& {Hummels}, C. 2020, \mnras, 492,
  1841, \dodoi{10.1093/mnras/stz3567}

\bibitem[{{McCourt} {et~al.}(2012){McCourt}, {Sharma}, {Quataert}, \&
  {Parrish}}]{McCourt2012}
{McCourt}, M., {Sharma}, P., {Quataert}, E., \& {Parrish}, I.~J. 2012, \mnras,
  419, 3319, \dodoi{10.1111/j.1365-2966.2011.19972.x}

\bibitem[{{McNamara} {et~al.}(2016){McNamara}, {Russell}, {Nulsen}, {Hogan},
  {Fabian}, {Pulido}, \& {Edge}}]{McNamara2016}
{McNamara}, B.~R., {Russell}, H.~R., {Nulsen}, P.~E.~J., {et~al.} 2016, \apj,
  830, 79, \dodoi{10.3847/0004-637X/830/2/79}

\bibitem[{{Mittal} {et~al.}(2012){Mittal}, {Oonk}, {Ferland}, {Edge}, {O'Dea},
  {Baum}, {Whelan}, {Johnstone}, {Combes}, {Salom{\'e}}, {Fabian}, {Tremblay},
  {Donahue}, \& {Russell}}]{Mittal2012}
{Mittal}, R., {Oonk}, J.~B.~R., {Ferland}, G.~J., {et~al.} 2012, \mnras, 426,
  2957, \dodoi{10.1111/j.1365-2966.2012.21891.x}

\bibitem[{{Owen} {et~al.}(2000){Owen}, {Eilek}, \& {Kassim}}]{Owen2000}
{Owen}, F.~N., {Eilek}, J.~A., \& {Kassim}, N.~E. 2000, \apj, 543, 611,
  \dodoi{10.1086/317151}

\bibitem[{{Qiu} {et~al.}(2019{\natexlab{a}}){Qiu}, {Bogdanovi{\'c}}, {Li}, \&
  {McDonald}}]{Qiu2019}
{Qiu}, Y., {Bogdanovi{\'c}}, T., {Li}, Y., \& {McDonald}, M.
  2019{\natexlab{a}}, \apjl, 872, L11, \dodoi{10.3847/2041-8213/ab0375}

\bibitem[{{Qiu} {et~al.}(2020){Qiu}, {Bogdanovi{\'c}}, {Li}, {McDonald}, \&
  {McNamara}}]{Qiu2020}
{Qiu}, Y., {Bogdanovi{\'c}}, T., {Li}, Y., {McDonald}, M., \& {McNamara}, B.~R.
  2020, Nature Astronomy, 4, 900, \dodoi{10.1038/s41550-020-1090-7}

\bibitem[{{Qiu} {et~al.}(2019{\natexlab{b}}){Qiu}, {Bogdanovi{\'c}}, {Li},
  {Park}, \& {Wise}}]{Qiu2018}
{Qiu}, Y., {Bogdanovi{\'c}}, T., {Li}, Y., {Park}, K., \& {Wise}, J.~H.
  2019{\natexlab{b}}, \apj, 877, 47, \dodoi{10.3847/1538-4357/ab18fd}

\bibitem[{{Qiu} {et~al.}(2021){Qiu}, {McNamara}, {Bogdanovic}, {Inayoshi}, \&
  {Ho}}]{Qiu2021}
{Qiu}, Y., {McNamara}, B.~R., {Bogdanovic}, T., {Inayoshi}, K., \& {Ho}, L.~C.
  2021, arXiv e-prints, arXiv:2103.06505.
\newblock \doarXiv{2103.06505}

\bibitem[{{Revaz} {et~al.}(2008){Revaz}, {Combes}, \& {Salom{\'e}}}]{Revaz2008}
{Revaz}, Y., {Combes}, F., \& {Salom{\'e}}, P. 2008, \aap, 477, L33,
  \dodoi{10.1051/0004-6361:20078915}

\bibitem[{{Russell} {et~al.}(2019){Russell}, {McNamara}, {Fabian}, {Nulsen},
  {Combes}, {Edge}, {Madar}, {Olivares}, {Salom{\'e}}, \&
  {Vantyghem}}]{Russell2019}
{Russell}, H.~R., {McNamara}, B.~R., {Fabian}, A.~C., {et~al.} 2019, \mnras,
  490, 3025, \dodoi{10.1093/mnras/stz2719}

\bibitem[{{Salom{\'e}} {et~al.}(2011){Salom{\'e}}, {Combes}, {Revaz}, {Downes},
  {Edge}, \& {Fabian}}]{Salome2011}
{Salom{\'e}}, P., {Combes}, F., {Revaz}, Y., {et~al.} 2011, \aap, 531, A85,
  \dodoi{10.1051/0004-6361/200811333}

\bibitem[{{Salom{\'e}} {et~al.}(2008){Salom{\'e}}, {Combes}, {Revaz}, {Edge},
  {Hatch}, {Fabian}, \& {Johnstone}}]{Salome2008}
---. 2008, \aap, 484, 317, \dodoi{10.1051/0004-6361:200809493}

\bibitem[{{Salom{\'e}} {et~al.}(2006){Salom{\'e}}, {Combes}, {Edge},
  {Crawford}, {Erlund}, {Fabian}, {Hatch}, {Johnstone}, {Sanders}, \&
  {Wilman}}]{Salome2006}
{Salom{\'e}}, P., {Combes}, F., {Edge}, A.~C., {et~al.} 2006, \aap, 454, 437,
  \dodoi{10.1051/0004-6361:20054745}

\bibitem[{{Scannapieco} \& {Br{\"u}ggen}(2015)}]{Scannapieco2015}
{Scannapieco}, E., \& {Br{\"u}ggen}, M. 2015, \apj, 805, 158,
  \dodoi{10.1088/0004-637X/805/2/158}

\bibitem[{{Schneider} {et~al.}(2018){Schneider}, {Robertson}, \&
  {Thompson}}]{Schneider2018}
{Schneider}, E.~E., {Robertson}, B.~E., \& {Thompson}, T.~A. 2018, \apj, 862,
  56, \dodoi{10.3847/1538-4357/aacce1}

\bibitem[{{Schure} {et~al.}(2009){Schure}, {Kosenko}, {Kaastra}, {Keppens}, \&
  {Vink}}]{Schure2009}
{Schure}, K.~M., {Kosenko}, D., {Kaastra}, J.~S., {Keppens}, R., \& {Vink}, J.
  2009, \aap, 508, 751, \dodoi{10.1051/0004-6361/200912495}

\bibitem[{{Sparre} {et~al.}(2019){Sparre}, {Pfrommer}, \&
  {Vogelsberger}}]{Sparre2019}
{Sparre}, M., {Pfrommer}, C., \& {Vogelsberger}, M. 2019, \mnras, 482, 5401,
  \dodoi{10.1093/mnras/sty3063}

\bibitem[{{Thompson} {et~al.}(2016){Thompson}, {Quataert}, {Zhang}, \&
  {Weinberg}}]{Thompson2016}
{Thompson}, T.~A., {Quataert}, E., {Zhang}, D., \& {Weinberg}, D.~H. 2016,
  \mnras, 455, 1830, \dodoi{10.1093/mnras/stv2428}

\bibitem[{{Truelove} {et~al.}(1997){Truelove}, {Klein}, {McKee}, {Holliman},
  {Howell}, \& {Greenough}}]{Truelove1997}
{Truelove}, J.~K., {Klein}, R.~I., {McKee}, C.~F., {et~al.} 1997, \apjl, 489,
  L179, \dodoi{10.1086/310975}

\bibitem[{{Veilleux} {et~al.}(2005){Veilleux}, {Cecil}, \&
  {Bland-Hawthorn}}]{Veilleux2005}
{Veilleux}, S., {Cecil}, G., \& {Bland-Hawthorn}, J. 2005, \araa, 43, 769,
  \dodoi{10.1146/annurev.astro.43.072103.150610}

\bibitem[{{Wang}(1995)}]{Wang1995}
{Wang}, B. 1995, \apj, 444, 590, \dodoi{10.1086/175633}

\bibitem[{{Zhang} {et~al.}(2017){Zhang}, {Thompson}, {Quataert}, \&
  {Murray}}]{Zhang2017}
{Zhang}, D., {Thompson}, T.~A., {Quataert}, E., \& {Murray}, N. 2017, \mnras,
  468, 4801, \dodoi{10.1093/mnras/stx822}

\bibitem[{{ZuHone} {et~al.}(2014){ZuHone}, {Biffi}, {Hallman}, {Randall},
  {Foster}, \& {Schmid}}]{ZuHone2014}
{ZuHone}, J.~A., {Biffi}, V., {Hallman}, E.~J., {et~al.} 2014, arXiv e-prints,
  arXiv:1407.1783.
\newblock \doarXiv{1407.1783}

\end{thebibliography}

\end{CJK*}

\end{document}